\documentclass{article}
\usepackage{spconf,amsmath,graphicx}

\usepackage{bbm}
\usepackage{algorithm}
\usepackage{algcompatible}
\usepackage{amssymb}
\usepackage{amsthm}
\newtheorem{theorem}{Theorem}
\usepackage{fancyhdr}

\title{High-dimensional confidence regions in sparse MRI}

\name{Frederik Hoppe$^{\star\circ}$\quad Felix Krahmer$^{\dagger}$\quad Claudio Mayrink Verdun$^{\dagger \textdollar\circ}$\quad Marion I. Menzel$^{\ddagger \S \mathparagraph}$\quad Holger Rauhut$^{\star}$\thanks{$^{\circ}$ Student contribution. Thanks to Federal Ministry of Education and Research in the grant \emph{``SparseMRI3D+ (FZK 05M20WOA)''}. M.I.M. acknowledges partial funding by the project AI for Health Imaging Award \emph{``CHAIMELEON: Accelerating the Lab to Market Transition of AI Tools for Cancer Management [H2020-SC1-FA-DTS-2019-1 952172]''}.}}
\address{$^{\star}$ Department of Mathematics, RWTH Aachen University, Aachen, Germany\\
$^{\dagger}$ Department of Mathematics, Technical University of Munich, Munich, Germany\\
$^{\ddagger}$ AImotion Bavaria, Technische Hochschule Ingolstadt, Ingolstadt, Germany\\
$^{\S}$ Department of Physics, Technical University of Munich, Garching, Germany\\
$^{\mathparagraph}$ GE Healthcare, Munich, Germany \\
$^{\textdollar}$ Chair of Theoretical Information Technology, Technical University of Munich, Munich, Germany\\
}

\begin{document}

\fancypagestyle{copyright}{\fancyhf{}\renewcommand{\headrulewidth}{0pt}\fancyfoot[L]{\small \copyright 2023 IEEE. Personal use of this material is permitted. Permission from IEEE must be obtained for all other uses in any current or future media, including reprinting/republishing this material for advertising or promotional purposes, creating new collective works, for resale or redistribution to servers or lists, or reuse of any copyrighted component of this work in other works.}}
\thispagestyle{copyright}

\maketitle

\begin{abstract}
One of the most promising solutions for uncertainty quantification in high-dimensional statistics is the debiased LASSO that relies on unconstrained $\ell_1$-minimization. The initial works focused on real Gaussian designs as a toy model for this problem. However, in medical imaging applications, such as compressive sensing for MRI, the measurement system is represented by a (subsampled) complex Fourier matrix. The purpose of this work is to extend the method to the MRI case in order to construct confidence intervals for each pixel of an MR image. We show that a sufficient amount of data is $n \gtrsim \max\{ s_0\log^2 s_0\log p, s_0 \log^2 p \}$.
\end{abstract}
\section{Introduction}
\label{sec:intro}

Several highly efficient methods for dealing with high-dimensional data have been proposed in recent decades. The idea of these methods is that the information contained in many natural datasets relies on statistics of much lower dimensions than the original ambient one. This innovation in statistics, signal processing, and machine learning became known as \emph{sparse regression} (SR) \cite{wainwright2019high,tibshirani1996regression} in the statistical literature or \emph{compressive sensing} (CS) \cite{donoho2006compressed, Foucart.2013} in the signal processing literature. However, a framework that quantifies uncertainty for guiding decision-making in certain critical applications, such as Magnetic Resonance Imaging (MRI), is still missing. Since reliable medical imaging procedures are pivotal for accurate interpretation and diagnostic tasks, a theory that quantifies the quality of such images based on sparse regression is highly relevant.

Recently, a series of papers initiated a de-sparsified approach to sparse regression \cite{Zhang.2014, Javanmard.2014, Javanmard.2018, vandeGeer.2014}. This technique can characterize the distribution of a modified estimator based on the KKT conditions of the LASSO solution. For this modified estimator, sharp confidence intervals are derived for variable selection in the case of (sub-)Gaussian designs.
Although these results provide fundamental theoretical insight for the uncertainty quantification theory in the high-dimensional regime, such fully random matrices are of limited practical use. In MRI, for example, the measurement process is highly structured and can be described by a (subsampled) Fourier operator \cite{rudelson2008sparse}. Moreover, structured matrices often allow for faster algorithmic processing by exploiting the fast Fourier transform (FFT) for matrix multiplication and efficient storage. The aim of this work is to close this gap by developing the theory for sharp confidence intervals for subsampled Fourier matrices since they are employed in the MR reconstruction pipeline.

\section{Background and related works}
Before we present our method for constructing confidence intervals in MRI, we introduce the underlying theory.

\subsection{Sparse regression}
For a design/measurement matrix $X \in \mathbb{C}^{n \times p}$ with rows $ x_1^T, \dots, x_n^T$ and a data vector $ y = (y_1, \dots y_n) \in \mathbb{C}^{n} $, we are interested in the high dimensional regression model

\begin{equation}\label{Cmodel}
y=X\beta^0+\varepsilon, \qquad p\gg n,
\end{equation}
where $\beta^0\in\mathbb{C}^p$ is $s_0$-sparse and the noise vector $\varepsilon\sim\mathcal{CN}(0,\sigma^2 I_{n\times n})$ is assumed to be a complex standard Gaussian vector whose components $\varepsilon_i$ are independent. Note that we are considering complex-valued representations since MRI measurements are typically modeled via complex numbers \cite{zhi2000principles}.

The main goal is to estimate $\beta^0 \in \mathbb{C}^p$ as well as to provide confidence regions for $\beta^0$ based on this estimator. 
A natural estimator is the LASSO \cite{tibshirani1996regression, maleki2013asymptotic}, denoted by $\hat{\beta}$, which is the minimizer of
\begin{equation}\label{eq:LASSO}
    \min\limits_{\beta\in\mathbb{C}^p}\frac{1}{2n}\Vert X\beta-y\Vert_2^2+\lambda\Vert \beta\Vert_1,
\end{equation}
where $\lambda=\lambda(n,p,\sigma) \in \mathbb{R}$ is a tuning parameter to balance the data fidelity term and sparsity induced by the $\ell_1$-norm. To stress the fact that all parameters are complex-valued, this problem is often referred to as the complex LASSO (c-LASSO) \cite{maleki2013asymptotic}.

\subsection{The desparsified LASSO}

Following the works \cite{Javanmard.2014, Javanmard.2018, vandeGeer.2014}, we aim to derive confidence bounds for the c-LASSO estimator in the case that the design matrix is given by a random subsampled Fourier matrix, discussed in more detail in Section \ref{sec:BOS}. Note that this is a matrix with heavy-tailed rows where standard concentration techniques cannot be trivially applied \cite{Vershynin.2018}. Most previous contributions assumed the design to be a real (sub-)Gaussian matrix $X\in\mathbb{R}^{n\times p}$, i.e. a matrix with light tailed distribution \cite{Javanmard.2014, Javanmard.2018, vandeGeer.2014}; only the work \cite{vandeGeer.2014} also provides results for fixed (deterministic) designs and bounded random designs under strong assumptions.

The \emph{debiased Lasso} estimator is constructed by "inverting" the KKT conditions \cite{vandeGeer.2014} and is defined as

\begin{equation}\label{eq:debiased_lasso}
    \hat{\beta}^u= \hat{\beta}+\frac{1}{n}MX^*(y-X\hat{\beta}),
\end{equation}
where $M$ may be chosen such that $M \hat{\Sigma} \approx I_{p\times p}$, where $\hat{\Sigma}$ is given by the sample covariance matrix, i.e., $\hat{\Sigma}=X^*X/n$. The difference between the debiased LASSO and the ground truth can then be decomposed into

\begin{equation}
    \sqrt{n}(\hat{\beta}^u - \beta^0) = \frac{M X^* \varepsilon}{\sqrt{n}} - \sqrt{n}(M \hat{\Sigma} - I_{p\times p})(\hat{\beta} - \beta^0).
\end{equation}
One of the important achievements of the desparsified LASSO theory is that it can be shown that the bias term $ R:=(M \hat{\Sigma} - I_{p\times p})(\hat{\beta} - \beta^0) $ asymptotically vanishes \cite{vandeGeer.2014}. As $ M X^* \varepsilon /\sqrt{n} \sim\mathcal{N}(0,\sigma^2\hat{\Sigma}) $, this allows us to construct pointwise confidence intervals for $\beta^0$.

The previous approaches estimate the terms $ (M \hat{\Sigma} - I_{p\times p})$ and $ (\hat{\beta} - \beta^0) $ separately which leads to non-optimal bounds and therefore to a sample size $n \gtrsim s_0^2 \log^2 p$ \footnote{Here, the notation $a \gtrsim b$ means that there is a constant $C>0$ such that $a \geq Cb$.}. The only exception requiring a sample size $n \gtrsim s_0 \log^2 p$ is the seminal paper \cite{Javanmard.2018} that, unlike the other works, uses a leave-one-out argument and strongly exploits the independence of $X\Sigma^{-1}e_i$ and $X_{-i}$. This independency holds, for example, for matrices with Gaussian rows but does not hold for heavy-tailed matrices such as a subsampled Fourier matrix.

\subsection{Subsampled Fourier matrices}\label{sec:BOS}

As a result of the Bloch equation, the magnetic resonance (MR) phenomenon can be modeled by Fourier measurements \cite{zhi2000principles}. A Fourier matrix $F\in\mathbb{C}^{p\times p}$ is defined entrywise as $F_{l,k}=e^{2\pi i(l-1)(k-1)/p}$ with $l,k\in[p]$. The random subsampled Fourier matrix $F_{\Omega}$, which plays a crucial role in fast MR image reconstruction, consists of the rows whose indices $j\in\Omega$ are obtained by $n$ independently and uniformly selected points from $[p]$. We denote these rows by $f_1^T,\hdots, f_p^T$. Note that with a probability larger than $0$, some indices may be chosen more than once. In practice, however, the rows are sampled without replacement. The results differ only slightly, as discussed in \cite[Chapter 12.6]{Foucart.2013}. Due to the sampling pattern, the rows of $F_{\Omega}$ are independent, but the entries within each row are not independent.

This type of measurement matrix falls into the class of bounded orthonormal systems \cite{Foucart.2013}. To simplify the exposition, here we state the results for subsampled Fourier matrices but note that they also hold for general matrices associated to bounded orthonormal systems as established in \cite{journal2022}.

\section{Confidence regions in the subsampled Fourier case}

As a consequence of the sampling pattern, the second-moment matrix $\mathbb{E}[f_jf_j^*]$ of any row $f_j$ is the identity $I_{p\times p}$ \cite{journal2022}. In addition, a subsampled Fourier matrix satisfies $\mathbb{E}\big[\hat{\Sigma}\big]=I_{p\times p}.$ Even though this is only in expectation, the estimation $\vert\hat{\Sigma}_{ij}\vert\leq 1$ for $i,j\in[p]$ shows that the entries of the sample covariance are restricted to the range $[0,1]$. Therefore, we choose $M=I_{p\times p}$. Then, the debiased LASSO from \eqref{eq:debiased_lasso} takes the form
\begin{equation}
    \hat{\beta}^u=\hat{\beta}+\frac{F_{\Omega}^*(y-F_{\Omega}\hat{\beta})}{n}.
\end{equation}

Our main theoretical result, Theorem \ref{thm:main_theoretical_result}, states that conditioned on $F_{\Omega}$, the debiased LASSO estimator is asymptotically normal, i.e.

\begin{equation}\label{eq:asym_normal}
    \sqrt{n}(\beta^u-\beta^0)\mid F_{\Omega}\sim\mathcal{CN}(0,\sigma^2\hat{\Sigma}).
\end{equation}
The crucial point for this asymptotic normality is that the bias term $R$ vanishes. This is the case if
\begin{equation}
    n \gtrsim \max\{ s_0\log^2 s_0\log p, s_0 \log^2 p \} ,
\end{equation}
which is further discussed in Section \ref{theory}.
Then, for a consistent noise estimator $\hat{\sigma}$ (see Section \ref{subsec:noise_estimation}), the confidence regions with significance level $\alpha\in(0,1)$ for $\beta_i^0\in\mathbb{C}$, estimated via the debiased LASSO,
\begin{equation}\label{eq:conf_region}
    J^{\circ}_i(\alpha):=\{z\in\mathbb{C}:\vert\hat{\beta}_i^u-z\vert\leq\delta_i^{\circ}(\alpha)\},
\end{equation}
with radius $\delta^{\circ}(\alpha):=\frac{\hat{\sigma}}{\sqrt{n}}\sqrt{\log(1/\alpha)}$
are asymptotically valid:
\begin{equation}
    \lim\limits_{n\to\infty}\mathbb{P}\left(\beta_i^0\in J^{\circ}_i(\alpha)\right)=1-\alpha.
\end{equation}

The results in \cite{Cai.2017} and \cite{Javanmard.2018} prove the optimality of the length of this type of confidence interval construction. In particular, they show that the optimal radius should scale with $\frac{1}{\sqrt{n}}$. In this sense, our confidence regions are optimal, and their construction follows straightforwardly from the asymptotic normality. A more detailed discussion on this procedure in the complex case can be found in \cite{journal2022}. 

\begin{algorithm}
\caption{Confidence regions in Fourier case}\label{alg:cap}
\begin{algorithmic}
\STATE Initialize $\alpha$, $F_{\Omega}$
\STATE Estimate noise $\hat{\sigma}$
\STATE $\lambda\gets$ cross validation: test multiples of $\lambda_0=\frac{\sigma\sqrt{K}}{\sqrt{n}}(2+\sqrt{12\log p})$
\STATE Solve LASSO $\hat{\beta}\gets \arg\min_{\beta}\frac{1}{2n}\Vert F_{\Omega}\beta-y\Vert_2^2+\lambda\Vert\beta\Vert_1$
\STATE Compute $\hat{\beta}^u\gets \hat{\beta}+F_{\Omega}^*(y-F_{\Omega}\hat{\beta})/n$
\STATE Compute $\delta^{\circ}(\alpha)\gets \frac{\hat{\sigma}}{\sqrt{n}}\sqrt{\log(1/\alpha)}$
\end{algorithmic}
\end{algorithm}

\section{Asymptotic normality of debiased LASSO in the subsampled Fourier case}
\label{theory}
The asymptotic normality mentioned in \eqref{eq:asym_normal} is the key property of the debiased LASSO to construct confidence intervals. It is stated rigorously in Theorem \ref{thm:main_theoretical_result}, which relies on the $\ell_2$-consistency of the LASSO estimator.

\subsection{$\ell_2$-consistency of LASSO}
One of the main sufficient conditions on the measurement matrix for establishing consistency of the LASSO estimator and optimal oracle inequalities \cite{SaraA.vandeGeer.2009} is the following: A matrix $X$ satisfies the restricted isometry property (RIP) of order $1\leq s\leq p$ with constant $\delta_s\in(0,1)$ if
\begin{equation}
    (1-\delta_s)\Vert \beta\Vert_2^2\leq \Vert X\beta\Vert_2^2\leq (1+\delta_s)\Vert \beta\Vert_2^2
\end{equation}
for all $s$-sparse vectors $\beta\in \mathbb{C}^p$. Although not being described by a light-tailed probabilistic model, (normalized) subsampled Fourier matrices still act as quasi-isometries on the subset of sparse vectors, i.e., they satisfy the RIP of order $s_0$ with high probability provided that $n\gtrsim s_0 \log^2 s_0 \log p$ \cite{Haviv.2017}.

This property can be used to show one of the key tools for asymptotic normality of desparsified estimators, namely, the existence of sharp $\ell_1$ and $\ell_2$ oracle estimates. In order to establish a bound for the bias term $R=(M \hat{\Sigma} - I_{p\times p})(\hat{\beta} - \beta^0)$, we start by stating oracle bounds for $\hat{\beta} - \beta^0$:
\begin{equation}
    \Vert \hat{\beta}-\beta^0\Vert_2\lesssim \frac{\sqrt{s_0\log p}}{\sqrt{n}}, \qquad  \Vert\hat{\beta}-\beta^0\Vert_1\lesssim \frac{ s_0\sqrt{\log p}}{\sqrt{n}}.
\end{equation}
These results are widely available in the statistics literature, where it is usually assumed that the design matrix fulfills the restricted eigenvalue condition \cite[Chapter 7]{wainwright2019high} or the compatibility condition \cite[Chapter 6]{Buhlmann.2011}. As it is standard in the compressive sensing literature, the measurement matrix here is assumed to satisfy the (slightly stronger) RIP \cite{Foucart.2013}. See \cite{SaraA.vandeGeer.2009} for a discussion about the different sufficient conditions for sparse regression and the relationship between them.

\subsection{Main theoretical result}\label{sec:asymptotic_normal}

\begin{theorem}\label{thm:main_theoretical_result}\cite{journal2022}
Let $\frac{1}{\sqrt{n}}F_{\Omega}$ be a normalized subsampled Fourier matrix with $n \gtrsim s_0\log^2 s_0\log p$ rows. Let further $\lambda\geq 2\lambda_0:= 2\frac{\sigma\sqrt{K}}{\sqrt{n}}(2+\sqrt{12\log p})$. Then, the following decomposition holds
\begin{equation}
    \sqrt{n}(\hat{\beta}^u-\beta^0)=W+R,
\end{equation}
where the debiased LASSO $\hat{\beta}^u$ is defined in \eqref{eq:debiased_lasso} and $W\mid F_{\Omega}\sim\mathcal{N}(0,\sigma^2\hat{\Sigma})$. Furthermore,
\begin{equation}\label{eq:main_result}
    \mathbb{P}\left(\Vert R\Vert_{\infty}\geq C(\sigma,\delta_t)\frac{\sqrt{s_0}\log p}{\sqrt{n}}\right)\leq 5p^{-2}
\end{equation}
with $C(\sigma,\delta_t)\geq 0$ depending only on $\sigma$ and $\delta_t<1$.
\end{theorem}

In order to guarantee that the bias term $R$ vanishes and hence, $\sqrt{n}(\hat{\beta}^u-\beta^0)$ is asymptotically Gaussian distributed, two sufficient conditions play a role - the fact that the measurement matrix satisfies the RIP with constant $\delta_t$, which requires $n \gtrsim  s_0\log^2 s_0\log p$ samples \cite{Haviv.2017} and the fact that the bias term $R$ asymptotically vanishes if $n \gtrsim  s_0\log^2 p $, as stated in \eqref{eq:main_result}. Therefore, in very precise terms, our sample complexity reads as $ n \gtrsim \max\{ s_0\log^2 s_0\log p, s_0\log^2 p \}$.

\subsection{Noise estimation}\label{subsec:noise_estimation}

From the theoretical point of view, estimating the error variance for high-dimensional estimators is a non-trivial problem. The most common method used in the debiased LASSO literature, e.g., in \cite{Zhang.2014, bellec2022biasing, Javanmard.2018, Li.2020, vandeGeer.2014} is the so-called scaled LASSO \cite{sun2012scaled}. From the MRI practitioners' point of view, the noise can be measured directly during MR image acquisition (the so-called pre-scan procedure, which is mandatory for every patient), yielding a direct, ground truth estimation of the noise \cite{AjaFernandez.2016}. Alternatively, it may be estimated retrospectively (indirectly) from the final image if the directly determined noise estimation is no longer accessible. For a review of noise estimation methods, see, e.g., \cite{AjaFernandez.2016}. In any case, a noise estimator is important for constructing confidence intervals. Lemma 13 in \cite{Javanmard.2014} shows that this is exactly the case, i.e., they show that the asymptotic normality still holds when the true noise level is replaced by a consistent noise estimator.

\section{Numerical Experiments}
\label{numerics}
\begin{figure}[t]
\begin{minipage}[b]{1.0\linewidth}
  \centering
  \centerline{\includegraphics[width=7.5cm]{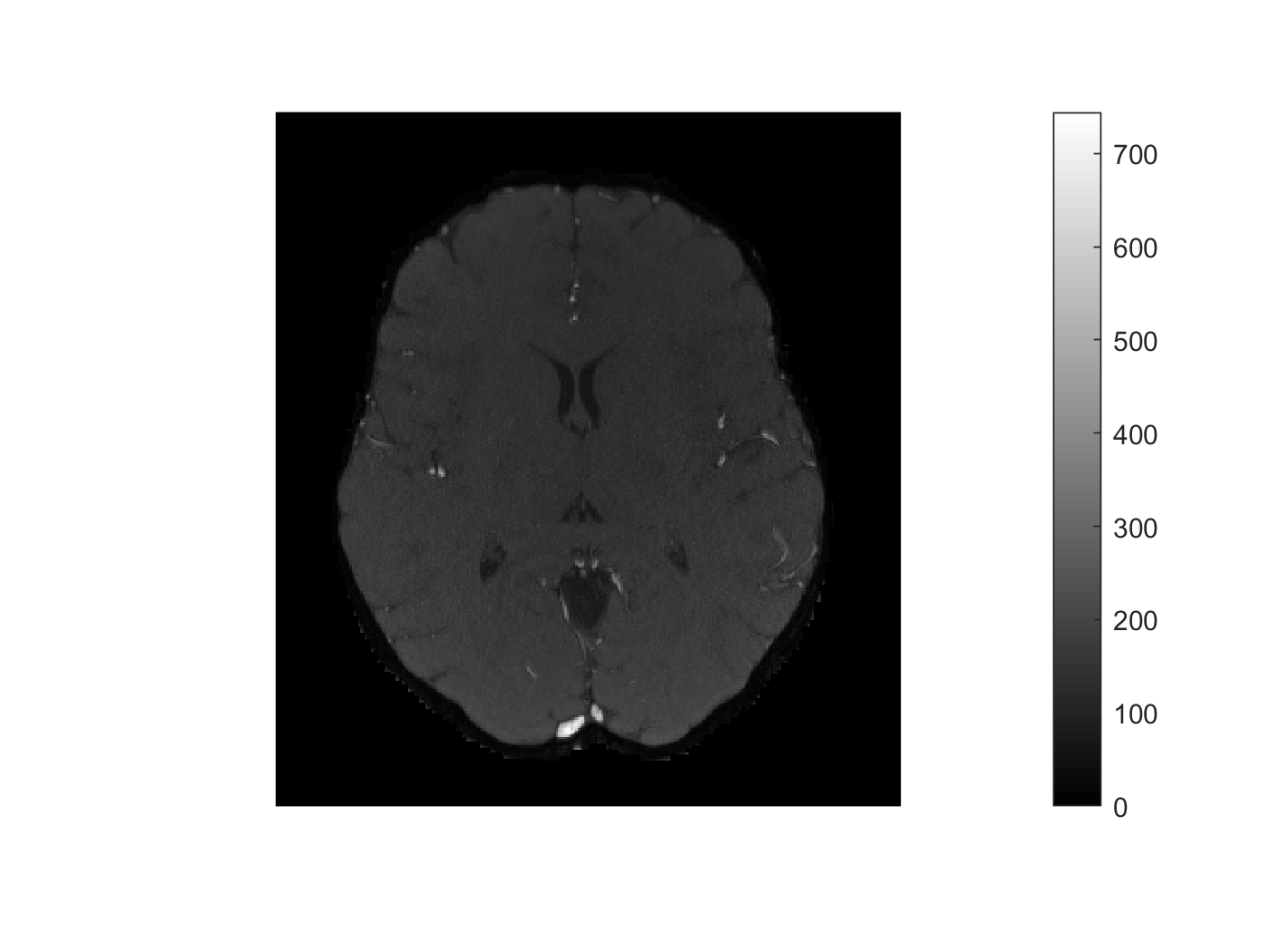}}
\end{minipage}
\caption{Original non-sparse MR angiography; single brain slice with vessels lighting up; image intensities in arbitrary units [u.a.].}
\label{fig:non_sparse_angio}
\end{figure}

In this section, we illustrate our theoretical results with numerical experiments using angiography brain image data from the Brain Vasculature (BraVa) database \cite{Wright.2013}, which is sparse in the canonical basis \cite{lustig}, as depicted in Figure \ref{fig:non_sparse_angio}. We use TFOCS \cite{becker2011templates} for the c-LASSO estimator in Algorithm \ref{alg:cap}. Throughout our experiments, we assume for simplicity that $\sigma$ is known and set $\alpha=0.05$.

\begin{figure}[t]
\begin{minipage}[b]{1.0\linewidth}
  \centering
  \centerline{\includegraphics[width=6.8cm]{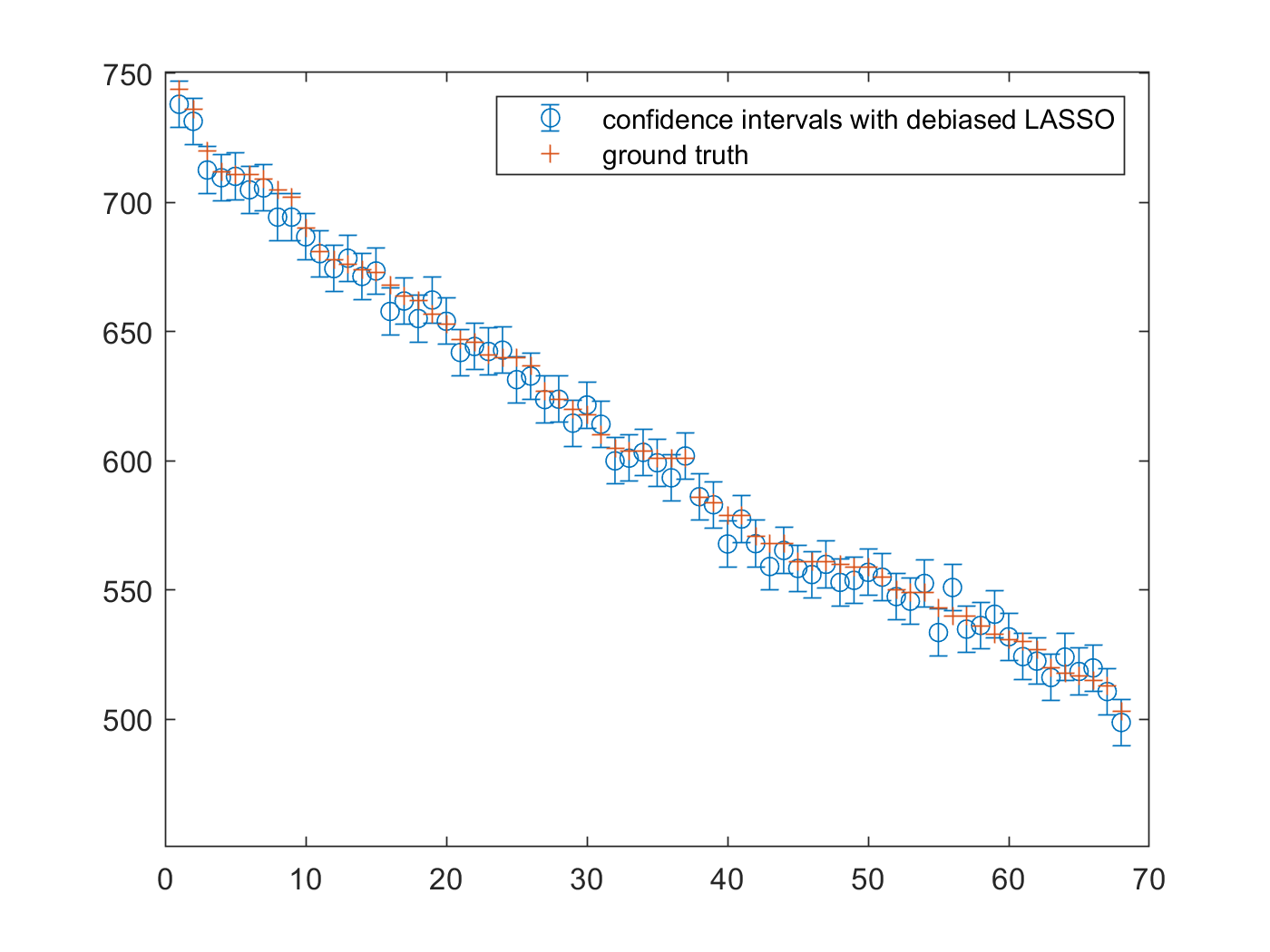}}
\end{minipage}
\caption{Confidence intervals based on the debiased LASSO for the pixels with the largest magnitude sorted in descending order.}
\label{fig:conf_int_largest_pixels}
\end{figure}

To artificially increase the sparsity of the input angiography image, highlighting only vessels, we set an image intensity threshold of 200 [a.u.], removing the brain background image and noise floor, such that each pixel with a magnitude lower than this threshold is set to $0$. We obtain an $s_0=1282$-sparse image, which serves as the (unknown) ground truth $\beta^0\in\mathbb{R}^{92160}$. Note that the theory, as well as the algorithm, allow for complex images. Still, due to better visualization, we stick to the real case where we simply set the imaginary part equal to $0$.
In the following, we simulate the image acquisition process in MRI by subsampling $n=0.4p=36864$ different rows of a full Fourier matrix. Then, we add complex noise with $\sigma=1000$ to the measurement vector $F_{\Omega}\beta^0$ leading to a relative noise level $\frac{\Vert\varepsilon\Vert_2}{\Vert F_{\Omega}\beta^0\Vert_2}=0.106$. From the model $y=F_{\Omega}\beta^0+\varepsilon$, we know $F_{\Omega}$, the subsampled data $y\in\mathbb{C}^{36864}$, and the noise level $\sigma$. The goal is to reconstruct the image $\beta^0$ and to provide lower and upper bounds for this estimate. Following Algorithm \ref{alg:cap}, with $\lambda=25\lambda_0$ chosen via cross-validation \cite{Chetverikov.07.05.2016}, we derive confidence regions. Figure \ref{fig:conf_int_largest_pixels} illustrates the confidence intervals based on the debiased LASSO estimator and the ground truth for the 68 pixels with the largest magnitudes (vessels). In order to measure the performance of our method, we define the hit rate and the hit rate on the support $S_0$, respectively, as 
\begin{equation}
    h=\frac{1}{p}\sum\limits_{i=1}^p\mathbbm{1}_{\{\beta^0_i\in J_i^{\circ}\}},\quad h_{S_0}=\frac{1}{s_0}\sum\limits_{i\in S}\mathbbm{1}_{\{\beta^0_i\in J_i^{\circ}\}}.
\end{equation}
We calculate the average hit rates for 100 realizations of the subsampled Fourier matrix and the noise. The results are presented in Table \ref{tab:hitrates}. Furthermore, we change the threshold leading to different sparsity levels in order to understand the role of the sparsity in the construction of the confidence intervals. Besides the hit rates, we calculate the similarity measure, SSIM, between the ground truth image and the estimated image. The smaller the sparsity, the better the hit rates and the SSIM. We observe the same behavior if we fix the sparsity and increase the amount of data $n$. Even though the (sufficient) condition $n\gtrsim s_0\log^2 p$ is not fulfilled for any threshold, the method still works well. Note that the hit rates do not depend on the noise level since the radius of confidence regions in \eqref{eq:conf_region} scales with the noise level. For example, a threshold of 200 and a noise level of $\sigma=2000$ lead to hit rates of, respectively, $h_{S_0}=0.932$ and $h=0.951$.

\begin{table}[t]
    \centering
    \begin{tabular}{c|c|c|c|c}
         threshold & $s_0$ & $h_{S_0}$ & $h$ & SSIM \\
         \hline 
         210 & 648 & 0.942 & 0.955 & 0.967 \\
         200 & 1282 & 0.931 & 0.951 & 0.964 \\
         190 & 2789 & 0.901 & 0.941 & 0.954 \\
         180 & 5510 & 0.823 & 0.916 & 0.889
    \end{tabular}
    \caption{Values of $h_{S_0}$, $h$ and SSIM for different sparsity levels and constant $n=0.4 p$. The values are averaged over 100 realizations of $F_{\Omega}$ and $\varepsilon$.}
    \label{tab:hitrates}
\end{table}

\section{Conclusion and future work}
\label{conclusion}

We derived confidence intervals for subsampled Fourier measurements that are used in compressive sensing for MRI reconstruction. The length of the confidence intervals decreases with the optimal rate $\frac{1}{\sqrt{n}}$. We showed that a sufficient amount of data for performing uncertainty quantification is given by $n \gtrsim \max\{ s_0\log^2 s_0\log p, s_0 \log^2 p \}$. For this purpose, we debiased the LASSO and showed its asymptotic normality. As an extension, we plan to derive confidence regions for quantitative multi-parametric MRI problems as well as for the case where the ground truth image is not trivially sparse but rather needs to be sparsified with a learned dictionary.

\vfill\pagebreak
\bibliographystyle{IEEEbib}
\bibliography{strings,refs}

\begin{thebibliography}{10}

\bibitem{wainwright2019high}
M.~J. Wainwright,
\newblock {\em High-dimensional statistics: A non-asymptotic viewpoint},
  vol.~48,
\newblock Cambridge University Press, 2019.

\bibitem{tibshirani1996regression}
R.~Tibshirani,
\newblock ``Regression shrinkage and selection via the lasso,''
\newblock {\em Journal of the Royal Statistical Society: Series B (Statistical
  Methodology)}, vol. 58, no. 1, pp. 267--288, 1996.

\bibitem{donoho2006compressed}
D.~L. Donoho,
\newblock ``Compressed sensing,''
\newblock {\em IEEE Transactions on Information Theory}, vol. 52, no. 4, pp.
  1289--1306, 2006.

\bibitem{Foucart.2013}
S.~Foucart and H.~Rauhut,
\newblock {\em A Mathematical Introduction to Compressive Sensing},
\newblock {Springer New York}, New York, NY, 2013.

\bibitem{Zhang.2014}
C.-H. Zhang and S.~S. Zhang,
\newblock ``Confidence intervals for low dimensional parameters in high
  dimensional linear models,''
\newblock {\em Journal of the Royal Statistical Society: Series B (Statistical
  Methodology)}, vol. 76, no. 1, pp. 217--242, 2014.

\bibitem{Javanmard.2014}
A.~Javanmard and A.~Montanari,
\newblock ``Confidence intervals and hypothesis testing for high-dimensional
  regression,''
\newblock {\em Journal of Machine Learning Research}, vol. 15, pp. 2869--2909,
  2014.

\bibitem{Javanmard.2018}
A.~Javanmard and A.~Montanari,
\newblock ``Debiasing the lasso: Optimal sample size for {Gaussian} designs,''
\newblock {\em The Annals of Statistics}, vol. 46, no. 6A, 2018.

\bibitem{vandeGeer.2014}
S.~{van de Geer}, P.~B{\"u}hlmann, Y.~Ritov, and R.~Dezeure,
\newblock ``On asymptotically optimal confidence regions and tests for
  high-dimensional models,''
\newblock {\em The Annals of Statistics}, vol. 42, no. 3, 2014.

\bibitem{rudelson2008sparse}
M.~Rudelson and R.~Vershynin,
\newblock ``On sparse reconstruction from {Fourier} and {Gaussian}
  measurements,''
\newblock {\em Communications on Pure and Applied Mathematics}, vol. 61, no. 8,
  pp. 1025--1045, 2008.

\bibitem{zhi2000principles}
Z.-P. Liang and P.~C. Lauterbur,
\newblock {\em Principles of Magnetic Resonance Imaging: A Signal Processing
  Perspective},
\newblock SPIE Optical Engineering Press, 2000.

\bibitem{maleki2013asymptotic}
A.~Maleki, L.~Anitori, Z.~Yang, and R.~G. Baraniuk,
\newblock ``Asymptotic analysis of complex lasso via complex approximate
  message passing (camp),''
\newblock {\em IEEE Transactions on Information Theory}, vol. 59, no. 7, pp.
  4290--4308, 2013.

\bibitem{Vershynin.2018}
R.~Vershynin,
\newblock {\em High-Dimensional Probability: An Introduction with Applications
  in Data Science},
\newblock Cambridge University Press, 2018.

\bibitem{journal2022}
F.~Hoppe, F.~Krahmer, C.~{Mayrink Verdun}, {M. I.} Menzel, and H.~Rauhut,
\newblock ``Uncertainty quantification for sparse {Fourier} recovery,''
\newblock {\em arXiv:2212.14864}, 2022.

\bibitem{Cai.2017}
T.~T. Cai and Z.~Guo,
\newblock ``Confidence intervals for high-dimensional linear regression:
  Minimax rates and adaptivity,''
\newblock {\em The Annals of Statistics}, vol. 45, no. 2, 2017.

\bibitem{SaraA.vandeGeer.2009}
{S. A. van de Geer} and {P. B{\"u}hlmann},
\newblock ``On the conditions used to prove oracle results for the {LASSO},''
\newblock {\em Electronic Journal of Statistics}, vol. 3, pp. 1360--1392, 2009.

\bibitem{Haviv.2017}
I.~Haviv and O.~Regev,
\newblock ``The restricted isometry property of subsampled {Fourier}
  matrices,''
\newblock in {\em Geometric Aspects of Functional Analysis: Israel Seminar
  (GAFA) 2014--2016}, Bo'az Klartag and Emanuel Milman, Eds., pp. 163--179.
  {Springer International Publishing}, Cham, 2017.

\bibitem{Buhlmann.2011}
P.~B{\"u}hlmann and S.~{van de Geer},
\newblock {\em Statistics for High-Dimensional Data},
\newblock {Springer Berlin Heidelberg}, Berlin, Heidelberg, 2011.

\bibitem{bellec2022biasing}
P.~C. Bellec and C.-H. Zhang,
\newblock ``De-biasing the lasso with degrees-of-freedom adjustment,''
\newblock {\em Bernoulli}, vol. 28, no. 2, pp. 713--743, 2022.

\bibitem{Li.2020}
S.~Li,
\newblock ``Debiasing the debiased {LASSO} with bootstrap,''
\newblock {\em Electronic Journal of Statistics}, vol. 14, no. 1, 2020.

\bibitem{sun2012scaled}
T.~Sun and C.-H. Zhang,
\newblock ``Scaled sparse linear regression,''
\newblock {\em Biometrika}, vol. 99, no. 4, pp. 879--898, 2012.

\bibitem{AjaFernandez.2016}
S.~Aja-Fern{\'a}ndez and G.~Vegas-S{\'a}nchez-Ferrero,
\newblock {\em Statistical Analysis of Noise in MRI: Modeling, Filtering and
  Estimation},
\newblock Springer, Cham, 2016.

\bibitem{Wright.2013}
S.~N. Wright, P.~Kochunov, F.~Mut, M.~Bergamino, K.~M. Brown, J.~C. Mazziotta,
  A.~W. Toga, J.~R. Cebral, and G.~A. Ascoli,
\newblock ``Digital reconstruction and morphometric analysis of human brain
  arterial vasculature from magnetic resonance angiography,''
\newblock {\em NeuroImage}, vol. 82, pp. 170--181, 2013.

\bibitem{lustig}
M.~Lustig, D.~Donoho, and J.~M. Pauly,
\newblock ``{Sparse MRI: The application of compressed sensing for rapid MR
  imaging},''
\newblock {\em Magnetic Resonance in Medicine}, vol. 58, no. 6, pp. 1182--1195,
  2007.

\bibitem{becker2011templates}
S.~R. Becker, E.~J. Cand{\`e}s, and M.~C. Grant,
\newblock ``Templates for convex cone problems with applications to sparse
  signal recovery,''
\newblock {\em {Mathematical Programming Computation}}, vol. 3, no. 3, pp.
  165--218, 2011.

\bibitem{Chetverikov.07.05.2016}
D.~Chetverikov, Z.~Liao, and V.~Chernozhukov,
\newblock ``{On cross-validated LASSO in high dimensions},''
\newblock {\em The Annals of Statistics}, vol. 49, no. 3, pp. 1300 -- 1317,
  2021.

\end{thebibliography}

\end{document}